\documentclass[10pt,twocolumn]{article}

\usepackage[T1]{fontenc}
\usepackage{newtxtext,newtxmath}
\usepackage{microtype}
\usepackage{amsmath,bm}
\usepackage{graphicx}
\usepackage{geometry}
\usepackage{natbib}
\usepackage{caption}
\usepackage{titlesec}
\usepackage{fancyhdr}
\usepackage[hidelinks]{hyperref}

\titleformat{\section}{\normalfont\bfseries\fontsize{9.5}{11}\selectfont}{\thesection}{0.6em}{}
\titleformat{\subsection}{\normalfont\bfseries\fontsize{9}{10.5}\selectfont}{\thesubsection}{0.6em}{}
\titlespacing*{\section}{0pt}{10pt plus 2pt minus 1pt}{4pt}
\titlespacing*{\subsection}{0pt}{8pt plus 2pt minus 1pt}{3pt}
\setcitestyle{numbers,square,comma,sort&compress}

\newcommand{\paperTitle}{Actuator Disk Models reproduce Actuator Line Model power and thrust fluctuation statistics in wind farm simulations}
\newcommand{\paperAuthors}{Manuel Ayala \;|\; Dennice F. Gayme \;|\; Charles Meneveau}

\begin{document}

\twocolumn[
\begin{@twocolumnfalse}
\vspace*{-3mm}
{\small\bfseries SHORT COMMUNICATION\par}
\vspace{6mm}
{\fontsize{17}{20}\selectfont\bfseries \paperTitle\par}
\vspace{4mm}
{\fontsize{9.3}{11}\selectfont \paperAuthors\par}
\vspace{3mm}
{\fontsize{8.2}{9.7}\selectfont
Department of Mechanical Engineering, Johns Hopkins University, Maryland, USA\par
\textbf{Correspondence:} Manuel Ayala (\href{mailto:mayala5@jhu.edu}{mayala5@jhu.edu})\par
\textbf{Funding:} This research was funded by the National Science Foundation and the Department of Energy (via NSF grant CBET-2401013)\par
\textbf{Keywords:} power fluctuations | disk-averaged velocity | actuator-disk model | actuator-line model | large-eddy simulation | thrust fluctuations\par}
\vspace{3mm}
\noindent\fboxsep=5pt\colorbox{black!6}{\parbox{\dimexpr\textwidth-2\fboxsep\relax}{%
\fontsize{8.45}{10.1}\selectfont
\textbf{ABSTRACT}\par\vspace{1mm}
Actuator-disk models (ADM) are widely used in wind-farm simulations, but their accuracy in reproducing power and thrust temporal fluctuation statistics has not yet been compared to the more accurate and detailed predictions of actuator-line models (ALM). This work provides a detailed comparison of such model predictions under three distinct atmospheric conditions. Data is obtained from a database (JHTDB-Wind) containing actuator-line turbine-response and time-resolved flow field data from a large-eddy simulation of a small windfarm operating over a full diurnal cycle. Power and thrust time series corresponding to ADM are constructed from disk-averaged velocity signals and compared with the corresponding more detailed actuator-line signals over three temporal windows within the diurnal cycle. The analysis compares time series, power spectral densities, and both the fluctuation and increment probability density distributions. The results show that the ADM time-series capture the dominant temporal, spectral, and statistical features of the ALM values at resolved and disk-averaged flow-field time scales, i.e. slower than the rotor frequency. Effects of temporal filtering that are often applied to ADM inputs are examined. Overall, the results provide strong support for the use of ADM for predicting power and thrust fluctuation statistics in wind-farm simulations.}}
\vspace{4mm}
\end{@twocolumnfalse}
]
\fontsize{8.55}{10.15}\selectfont

\section{Introduction}\label{Intro}

Wind-turbine power and thrust fluctuate because turbines operate in turbulent atmospheric flows. In wind farms these quantities are also influenced by the wakes of upstream turbines. Power fluctuations are of great interest as they determine the unsteady electrical output of a turbine or wind farm and therefore affect grid integration, storage requirements, and control strategies \citep{apt2007,bandi2017,Shapiro2022TurbulenceControl,KatzensteinApt2012}. Thrust fluctuations determine the unsteady force exerted by the turbine on the flow and are directly related to wake dynamics, aerodynamic loading, fatigue, and turbine lifetime \citep{porteagel_review_2020,veers2023,Kosovic2026}. Accurately predicting both quantities is therefore important for wind-farm modeling, turbine-load assessment, as well as control-oriented design and operational studies.

Large-eddy simulations (LES) of wind farms require turbine representations that balance accuracy and cost. A relatively high-fidelity approach is the actuator-line model (ALM), in which rotating blades are represented by a large number of actuator points on a line along  the rotating blade span \citep{Shen2011MexicoRotor,MartinezTossas2015}. Lift and drag forces at each point are computed from local LES velocities and projected onto the LES grid \citep{MartinezTossas2015}, with appropriate models accounting for unresolved shed vorticity physics \citep{jha2018actuator,martinez2019filtered,daug2020new,martinez2024generalized}. The turbine power is obtained from the aerodynamic torque, $T(t)$, as
\begin{equation}
P(t)=T(t)\,\Omega(t),
\label{eq:alm_power}
\end{equation}
where $\Omega(t)$ is the rotor angular velocity. The ALM includes blade rotation, azimuthal sampling, and radial loading variations, and is therefore typically considered a   suitable model  for assessing power and thrust fluctuations from LES \citep{MartinezTossas2015,Liu2022ALMAccuracy}.

In order to reduce the computational burden, many wind-farm LES studies instead use the actuator-disk model (ADM), in which the turbine is represented as a permeable disk that extracts momentum and energy from the resolved flow \cite{Jimenez2010WakeDeflection}. ADM can typically be applied in LES with coarser meshes than ALM (see \cite{martinez2024generalized}, however).  The relevant inflow quantity for ADM is the disk-averaged velocity,
\begin{equation}
U_D(t) = \frac{1}{A}\int_{A_D} \boldsymbol{u}(\boldsymbol{x},t)\cdot \boldsymbol{n}\, \mathrm{d}A,
\label{eq:disk_average_velocity}
\end{equation}
where $\boldsymbol{u}(\boldsymbol{x},t)$ is the  fluid velocity resolved in LES. This velocity depends on the three dimensional position ${\bf x}$ and time $t$, $A_D$ is the swept area perpendicular to the airflow and $A=\pi D^2/4$ is the rotor area, and $\boldsymbol{n}$ is the rotor-normal direction. The corresponding ADM estimates of thrust and power are
\begin{equation}
F_{T}(t)=\frac{1}{2}\rho A C'_T U_D^2(t),
\label{eq:adm_thrust}
\end{equation}
and
\begin{equation}
P(t)=\frac{1}{2}\rho A C'_P U_D^3(t).
\label{eq:adm_power}
\end{equation}

Here, $C'_T$ and $C'_P$ are the ``local'' thrust and power coefficients defined with respect to the disk-averaged velocity \cite{meyers2010AIAA} rather than the nominally undisturbed upstream velocity. This formulation is attractive for wind-farm simulations because $U_D$ is directly available from the resolved flow, so it obviates the need for ambiguous definitions of upstream velocity for turbines operating in wakes~\citep{calaf2010large,meyers2010AIAA}.

Comparisons of wake and flow-field quantities, including mean wake profiles, turbulence intensity, grid-resolution effects, and force-projection choices \citep{PorteAgel2011LESFramework,MartinezTossas2015,Troldborg2015WakeProperties,Stevens2018ADMALMComparison} have shown that ADM and ALM can produce similar intermediate and far-wake flow statistics and, under controlled conditions, comparable mean power predictions.  Prior analyses of ADM-generated power fluctuations \cite{Stevens2014Temporal,Singh2024WindPowerFluctuationWakes,LiuStevens2026,Tobin2019SpatiotemporalCorrelations,Liu2024TurbulenceCoherenceTurbines,yang_offshorefarm_2014,yang2014large} assumed that the ADM generated realistic predictions of power and thrust fluctuations but this assumption has not been thoroughly tested. Therefore, it remains unclear whether ADM-based estimates can reproduce the temporal, spectral, and statistical structure of ALM power and thrust fluctuations. The purpose of this technical note is to describe detailed comparisons between ADM and ALM predictions of fluctuation statistics.

Using the JHTDB-Wind diurnal-cycle LES database, which provides both time-resolved flow fields and ALM turbine-response data, we construct ADM estimates of power and thrust using disk-averaged velocity. We compare these estimates with the corresponding ALM signals based on detailed blade-level time and position-dependent force information. Our analysis uses time series, power spectral densities, and probability density functions (PDFs) of fluctuations  and power fluctuation increments  to assess whether the ADM captures not only the mean response, but also the dominant fluctuation behavior of turbine power and thrust.

\begin{figure}[!t]
\centering
\includegraphics[width=\linewidth]{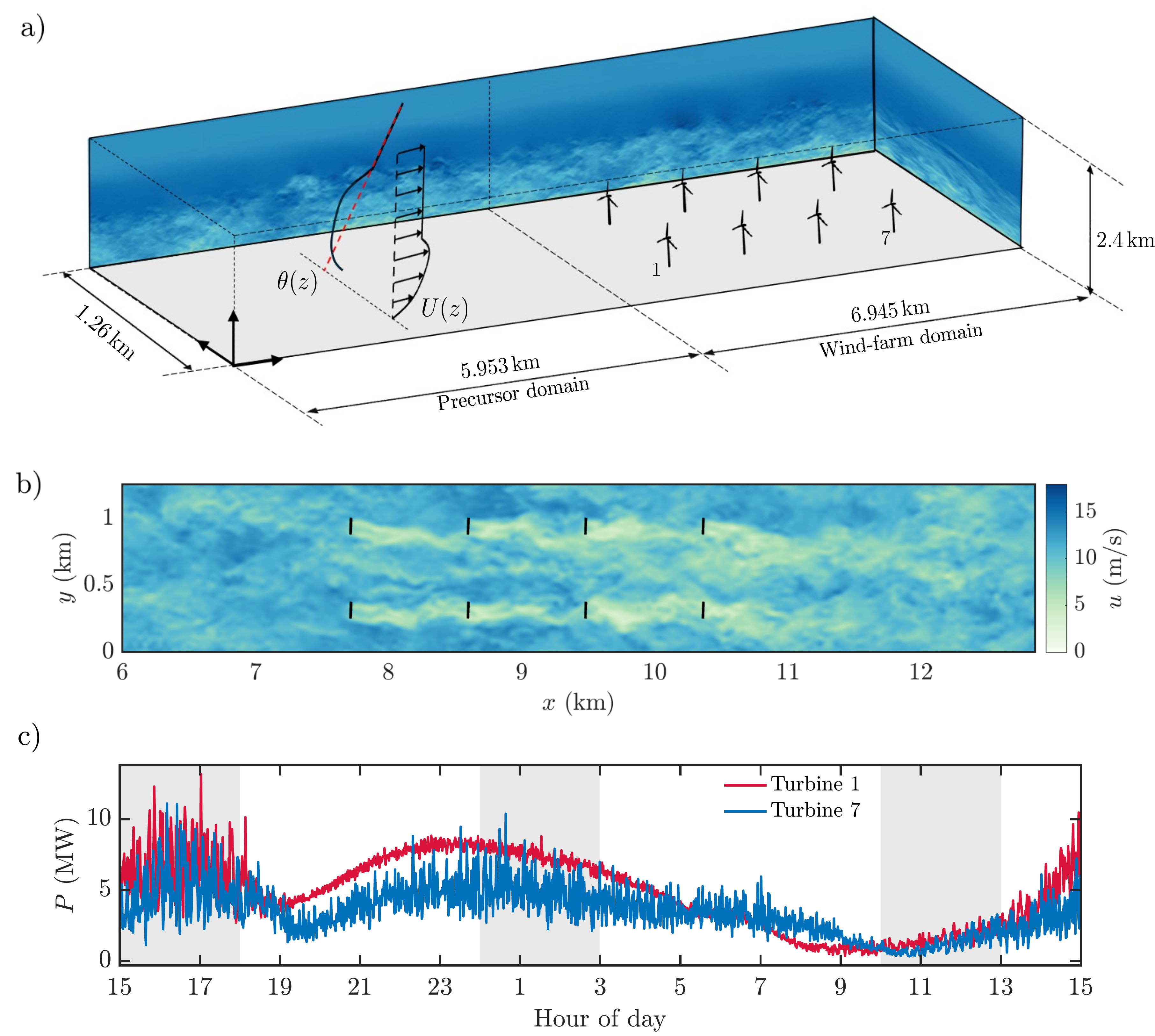}
\caption{a) Schematic of the domain in which flow data is stored during an entire 24 diurnal cycle. Color contours represent a snapshot of streamwise turbulent veloocty at 18:00. b)  Instantaneous contours of the streamwise velocity at hub height at 18:00, viewed from above. Black lines are the turbines. c) Turbine power signal over the full 24-h cycle (red line for a turbine in first row and blue for a waked turbine in last row); shaded regions indicate the three temporal windows considered in this study.}
\label{fig1}
\end{figure}

\section{JHTDB-Wind Diurnal Cycle Database}\label{database}

The present analysis uses the JHTDB-Wind small wind farm diurnal-cycle dataset described in detail in \cite{Xiao2025DiurnalJHTDBWind}. This dataset consists of both time-resolved flow fields and turbine-level ALM quantities from LES of an 8-turbine wind farm operating over a full 24-hour diurnal cycle. Figure~\ref{fig1} illustrates the database flow domain, an instantaneous hub-height flow field, and turbine power signals provided by the database. This dataset is well suited for the present study because the resolved velocity fields can be used to construct the ADM input variables, while the ALM turbine power and thrust provide the reference signals. The simulated wind farm contains 8 NREL-5MW+ turbines arranged in 4 streamwise rows and 2 spanwise columns, with streamwise and spanwise spacings of $7D$ and $5D$, respectively. The turbines have diameter $D=126$ m and hub height $z_h=90$ m. The flow is driven by a prescribed geostrophic wind, and the simulation includes Coriolis forcing, wind veer, and time-varying solar surface heating, leading to varying atmospheric stability that follows a typical diurnal cycle. During the ALM simulation, the turbines yaw to align with the mean hub-height inflow direction from the precursor domain.The turbines operate in below-rated, region-II conditions at an approximately optimal fixed tip-speed ratio at all times, so the rotor angular velocity varies in time with the local incoming velocity. 

For the present comparison, we use the stored three-dimensional velocity fields, available every $0.5$ s, to compute yaw-aligned disk-averaged velocity signals at the rotor plane. Turbine-level ALM data, including power, thrust, and rotor angular velocity, are available for each turbine at $\Delta t=0.05$ s. The ADM power and thrust estimates are then constructed from the disk-averaged velocity and compared directly with the corresponding ALM turbine-response signals. This setup enables a comparison of ADM and ALM power and thrust fluctuations across different temporal windows of the diurnal cycle.

\begin{figure*}[t!] 
\centerline{\includegraphics[width=\linewidth]{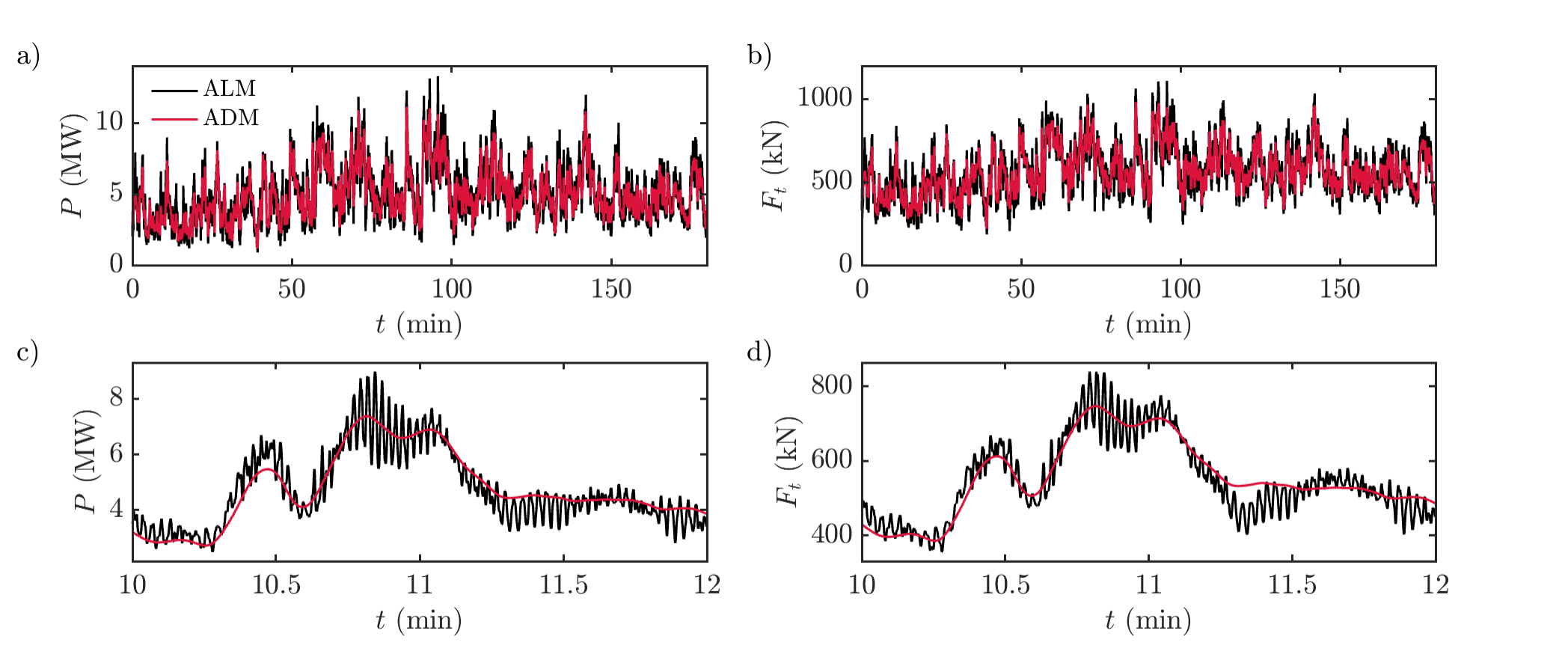}}\caption{Comparison of ALM and ADM turbine-response signals for turbine 7 during the 15:00--18:00 window. Panels (a) and (b) show the full time signals of power and thrust, respectively. Panels (c) and (d) show corresponding two-minute excerpts, highlighting the short-time-scale variability in the power and thrust signals.}
\label{fig2}
 
\end{figure*}

\section{Power and Thrust Fluctuations estimated by Disk-averaged Velocities}\label{results}

The analysis focuses  on turbine 7 (see Figure \ref{fig1}a), which is located in the last row and experiences wake-affected inflow. To evaluate the ADM performance under different atmospheric conditions, we consider three distinct temporal windows of the diurnal cycle (shown as shaded regions in Figure \ref{fig1}c): 15:00--18:00, representative of late-afternoon convective conditions; 00:00--03:00, representative of nocturnal stable conditions; and 10:00--13:00, representative of post-morning-transition daytime conditions. A single turbine-based  comparison allows us to  compare ADM and ALM fluctuation predictions  across these varying conditions without introducing the additional complexity of aggregating signals across multiple turbines. Since farm-level power and thrust are obtained from the superposition of individual turbine responses, verifying accurate recovery of the fluctuation behavior of representative single turbines is  sufficient  for evaluating the extent to which  ADM-based predictions match those of the more accurate ALM simulations.

To construct the ADM thrust and power predictions, we first compute the disk-averaged rotor-normal velocity from the time-resolved flow fields. The flow-field data used in this study are queried from JHTDB-wind with \texttt{getData(...)} at the yaw-rotated rotor-disk sampling locations and stored flow-field times. Since the turbines are yawed to ensure that the veer angle is zero at hub height, the rotor disk is not fixed in the $y$-$z$ plane. Instead, at each time $t$, the disk is rotated by the instantaneous turbine yaw angle $\gamma(t)$, and the averaged velocity is taken along the rotor-normal direction. The disk-averaged velocity is therefore computed as
\begin{equation}
U_D(t) =
\frac{1}{A}\int_{A_D(\gamma)}
\left[u(\boldsymbol{x},t)\cos\gamma(t)
+
v(\boldsymbol{x},t)\sin\gamma(t)\right]\,\mathrm{d}A,
\label{eq:disk_average_velocity_yaw}
\end{equation}
where $A_D(\gamma)$ the yaw-rotated swept area perpendicular to the airflow, and $u$ and $v$ are the streamwise and spanwise velocity components from LES data, respectively. In practice, the integral is evaluated by sampling the velocity over points contained within a circular disk of radius $D/2$ centered at the turbine hub (using spatial spline interpolation as afforded by JHTDB-wind querying capabilities) and averaging the rotor-normal velocity over those points at each stored flow-field time. Because the ALM rotor angular velocity $\Omega(t)$ is determined from the axial velocity sampled one grid cell ($\Delta=18.375$~m) upstream of the rotor, we evaluate the yaw-rotated disk-averaged velocity at the same location for consistency. We then apply a sharp low-pass spectral filter to the disk-averaged velocity, using the mean turbine rotational frequency in each temporal window as the cutoff. The cutoff frequencies are $f=0.203$ Hz for 15:00--18:00, $f=0.202$ Hz for 00:00--03:00, and $f=0.124$ Hz for 10:00--13:00. This filtering restricts the ADM predictions to variability on time scales longer than the rotor-rotation time scale, consistent with the rotor-disk-averaged nature of the ADM. The corresponding turbine-level ALM quantities, including yaw, power, thrust, and rotor angular velocity time-series, are obtained from JHTDB-wind using \texttt{getTurbineData(...)} calls \cite{Xiao2025DiurnalJHTDBWind}.

The ADM predictions of power and thrust are respectively computed using Eqs.~\eqref{eq:adm_power} and \eqref{eq:adm_thrust}. The coefficients $C'_P$ and $C'_T$ are first estimated independently for each temporal window of the diurnal cycle by requiring the mean ADM response to match the corresponding ALM response, with $C'_P=\overline{P_{\mathrm{L}}}/(0.5\rho A\,\overline{U_D^3})$ and $C'_T=\overline{F_{T,\mathrm{L}}}/(0.5\rho A\,\overline{U_D^2})$. Here, $\overline{P_{\mathrm{L}}}$ and $\overline{F_{T,\mathrm{L}}}$ are the time-averaged ALM power and thrust, respectively. The coefficients obtained for each region are then averaged to define a single pair of values used throughout the analysis. This procedure yields $C'_T=1.263$ and $C'_P=1.460$. These values are close to those predicted by classical actuator-disk momentum theory, for which $C'_T=C'_P=4a/(1-a)$; for a representative induction factor $a=0.25$, corresponding to $C'_T=C'_P\simeq1.33$ \cite{Stevens2014Temporal,meyers2010AIAA}.

\subsection{Power and Thrust Time Signals}

Figure~\ref{fig2} compares the ALM and ADM power and thrust time signals for turbine 7 during the 15:00--18:00 window. The ALM signals exhibit pronounced temporal variability, with fluctuations occurring both over the full window and over shorter time scales of a few minutes. The ADM estimates capture the dominant trends in both power and thrust, including the main low-frequency variations, while attenuating part of the short-time-scale variability that occurs on the time-scale of blade rotation (a few seconds). This attenuation is expected because the disk-averaging and temporal filtering reduces velocity fluctuations that are not coherent over the rotor area and eliminates the blade rotational flow sampling that is at higher frequency.

\subsection{Power Spectral Density of Power and Thrust Fluctuations}
We next compare the ALM and ADM power and thrust fluctuations using their power-spectra densities (PSDs). Fluctuations are defined by subtracting the temporal mean, e.g., $P'(t)=P(t)-\overline{P}$ and $F_T'(t)=F_T(t)-\overline{F_T}$. The PSDs are estimated using Welch's method within each three-hour window. Because the ALM and ADM signals are sampled at different rates, $\Delta t=0.05$ s and $\Delta t=0.5$ s, respectively, we use different segment lengths: $2^{14}$ samples for the ALM and $2^{11}$ samples for the ADM. Both estimates use 75\% overlap, a normalized Hanning window, and a Fourier-transform length equal to the segment length. These choices give comparable physical window durations for the ALM and ADM spectra, allowing comparison over the common range of resolved frequencies.

\begin{figure*}[t!]
\centerline{\includegraphics[width=\textwidth]{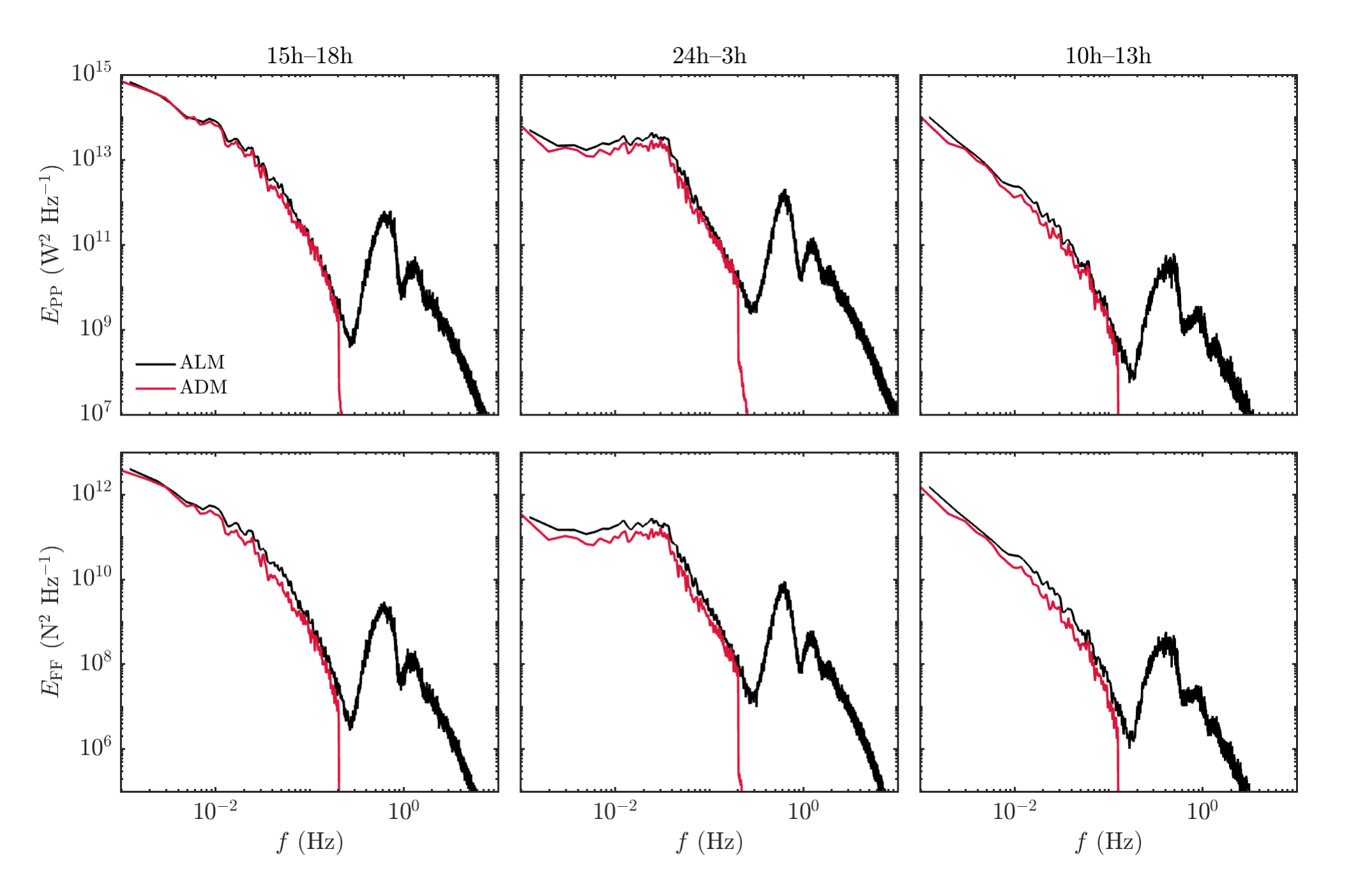}}
\caption{Power spectral densities of turbine \#7 power and thrust fluctuations for the ALM (black line) and ADM (red line) signals. The top row shows the PSDs of power fluctuations, $E_{PP}$, and the bottom row shows the PSDs of thrust fluctuations, $E_{FF}$. Columns correspond to the three temporal windows considered: 15:00--18:00, 00:00--03:00, and 10:00--13:00.}
\label{fig3}
\end{figure*}

As can be seen in Figure~\ref{fig3}, the PSDs obtained from the disk-averaged ADM estimates show good agreement with the ALM reference spectra for both power and thrust across the three temporal windows considered for frequencies below 0.25 Hz (i.e. time-scales longer than a few seconds). Over a broad range of resolved frequencies, the ADM reproduces the main spectral energy content and captures the stability-dependent changes in spectral decay observed in the ALM signals. The ALM spectra also contain distinct peaks and harmonics associated with the blade-passing frequency, approximately $f\approx 0.5$--$0.65$ Hz, which arise from the periodic sampling of the turbulent inflow by the rotating blades.  The overall spectral behaviors shown in Figure~\ref{fig3} vary substantially across the three temporal windows, reflecting changes in the turbulent inflow sampled by the turbines under different atmospheric stability conditions. In particular, differences in the spectral decay over the intermediate-frequency range suggest that the turbine response is influenced by the stability-dependent structure of the incoming turbulence.

\subsection{Effect of Temporal Filtering }

In actuator-disk simulations, the disk-averaged velocity used to compute turbine forcing is often temporally filtered after the spatial averaging over the rotor disk \cite{Stevens2018ADMALMComparison,Goit2015OptimalControl,Bempedelis2023TurbulentEntrainment}. This filtering is commonly motivated as a way to reduce the influence of high-frequency oscillations in the instantaneous disk-averaged velocity. A commonly used one-sided exponential relaxation filter is given by
\begin{equation}
U^T_D(t) = \epsilon U_D(t) + (1-\epsilon)U^T_D(t-\Delta t),
\label{eq:disk_average_velocity_temporalfilt}
\end{equation}
where $\epsilon=\Delta t/(\tau_f+\Delta t)$ is the relaxation coefficient, $\tau_f$ is the filter time scale, and $U^T_D$ is the temporally filtered disk-averaged velocity. The choice of $\tau_f$ varies widely across studies, from short values such as $\tau_f=0.2$ s \citep{Stevens2018ADMALMComparison} and $\tau_f=5$ s \citep{Allaerts_Meyers_2017,Goit2015OptimalControl} to values proportional to boundary-layer scales \citep{calaf2010large,Bempedelis2023TurbulentEntrainment}.

\begin{figure}[t!]
\centerline{\includegraphics[width=0.9\linewidth]{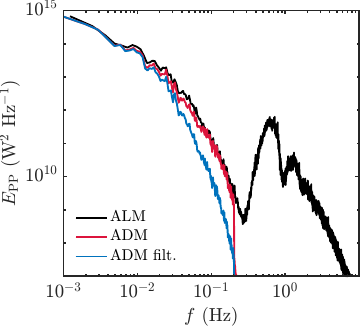}}
\caption{Effect of temporal filtering on the ADM spectral estimates for power production $E_{\rm PP}$ of turbine \#7 during the 15:00--18:00 window (similar results are obtained for the spectra of thrust).}
\label{fig4}
\end{figure}

Figure~\ref{fig4} shows the effect of applying this filter with $\tau_f=5$ s before computing the ADM response. As expected, the filtering modifies the spectral content, reducing the amplitude of the fluctuations and steepening the spectral decay at intermediate frequencies. Similar behavior is observed across all temporal windows. Thus, temporal filtering can significantly affect ADM-based fluctuation estimates, and the filter time scale should be chosen carefully.

\subsection{Probability Density Functions }

\begin{figure*}[ht!]
\centerline{\includegraphics[width=\linewidth]{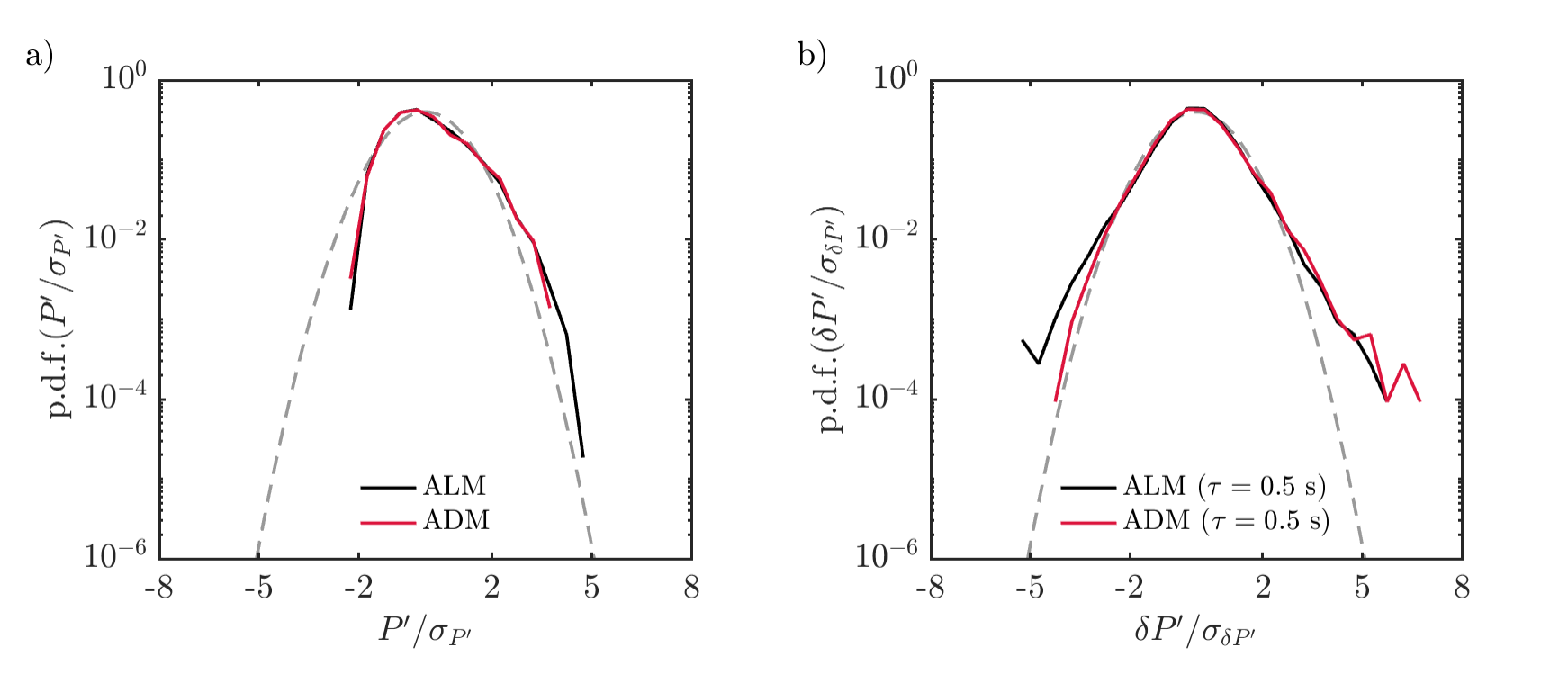}}
\caption{(a) Normalized probability density functions of power fluctuations for the ALM (black line) and ADM (red line) signals. (b) Probability density functions of normalized power fluctuation increments. The gray dashed line denotes a standard normal distribution with zero mean and unit variance.}
\label{fig5}
\end{figure*}

We further assess the ability of the ADM to reproduce the statistical structure of the ALM fluctuations by comparing probability density functions (PDFs) of the power fluctuations and power fluctuation increments. Figure~\ref{fig5}a shows the PDFs of the normalized power fluctuations. The ALM and ADM distributions agree well over the core of the PDFs, indicating that the ADM captures the most probable fluctuation amplitudes. Both distributions deviate from the Gaussian reference, reflecting the non-Gaussian character of the rotor-integrated turbine response. Differences between ADM and ALM are most apparent in the tails. However, these tail differences should be interpreted with caution because the two signals are sampled at different temporal resolutions: the ALM signal is available at $\Delta t=0.05$ s, whereas the ADM signal is available at $\Delta t=0.5$ s. Consequently, the ALM record contains more samples over the same temporal window and can better populate rare events in the far tails. Thus, the observed tail differences may reflect a combination of model-form effects, temporal-resolution effects, and finite-sample uncertainty, rather than disk averaging alone. Figures~\ref{fig5}b compares the PDFs of normalized fluctuation increments for power $\delta P'=P'(t)-P'(t-\tau)$. When the ALM increments are evaluated at $\tau=0.5$s, matching the sampling interval of the ADM signal, the ALM and ADM increment PDFs are quite close.  The PDFs of the thrust fluctuations and their increments exhibit behavior similar to that of the power fluctuations and are therefore omitted for brevity.

\section{Conclusions}\label{conclusions}

We conclude that the ADM captures the dominant temporal, spectral, and statistical features of the ALM response at the relevant resolved flow-field time scales. In the time domain, it reproduces the main low-frequency variability while attenuating part of the short-time-scale fluctuations due to rotor-disk averaging and blade rotation. In the frequency domain, the ADM spectra agree well with the ALM spectra over the resolved range and reproduce the changes in spectral behavior across the diurnal cycle. Differences are most evident at high frequencies where ADM does not contain frequencies associated with blade rotation. The PDF comparisons further show that the ADM reproduces the core of the normalized power and thrust fluctuation distributions. Overall, these results support the use of ADM-based estimates for power and thrust fluctuations in large wind-farm simulations, provided that their temporal-resolution and filtering characteristics are taken into account appropriately.

\section*{Acknowledgments}
This work was supported by the National Science Foundation and the Department of Energy (via NSF grant CBET-2401013).

%None reported.

%The authors declare no conflicts of interest.

\end{document}